\documentstyle[12pt]{article}

\newcounter{eq}
\newcounter{sc}

\def\overleftrightarrow#1{\vbox{\ialign{##\crcr
 $\leftrightarrow$\crcr\noalign{\kern-1pt\nointerlineskip}
 $\hfil\displaystyle{#1}\hfil$\crcr}}}

\newlength{\minitwocolumn}
\begin{document}

\begin{flushright}
DPUR/TH/56\\
June, 2017\\
\end{flushright}
\vspace{20pt}

\pagestyle{empty}
\baselineskip15pt

\begin{center}
{\large\bf Weinberg's No Go Theorem in Quantum Gravity
\vskip 1mm }

\vspace{10mm}
Ichiro Oda \footnote{E-mail address:\ ioda@phys.u-ryukyu.ac.jp
}

\vspace{3mm}
           Department of Physics, Faculty of Science, University of the 
           Ryukyus,\\
           Nishihara, Okinawa 903-0213, Japan.\\

\end{center}


\vspace{3mm}
\begin{abstract}
An important hurdle to be faced by any model proposing a resolution to the cosmological constant problem
is Weinberg's venerable no go theorem. This theorem states that no local field equations including classical
gravity can have a flat Minkowski solution for generic values of the parameters, in other words, the no go
theorem forbids the existence of any solution to the cosmological constant problem within local field theories
without fine tuning.  Though the original Weinberg theorem is valid only in classical gravity, in this article we prove
that this theorem holds even in quantum gravity. Our proof is very general since it makes use of the BRST 
invariance emerging after gauge-fixing of general coordinate invariance and  does not depend on the detail 
of quantum gravity. 
\end{abstract}

\newpage
\pagestyle{plain}
\pagenumbering{arabic}


\rm

One of the most interesting but difficult problems of modern theoretical physics and cosmology concerns
the cosmological constant, in particular, its smallness and its severe fine tuning. Most of the models addressing
the cosmological constant problem thus far rely on some dynamical mechanism where some classical field
configuration adjusts a bare vacuum energy density to a tiny value. Any proposal of this kind, however, is 
in conflict with a celebrated no go theorem due to Weinberg \cite{Weinberg}, which states that 
no local field equations including classical gravity can have a flat Minkowski solution for generic values of 
the parameters. To put differently, the no go theorem forbids the existence of any solution to the cosmological
constant problem within local field theories unless there is a fine tuning.

To bypass the Weinberg theorem, therefore, it seems that we should move on to some kinds of nonlocal
field theories. Actually, there have recently appeared such two classes of nonlocal models, those are, 
the vacuum energy sequester  \cite{Kaloper1, Kaloper2, Kaloper3, Kaloper4, Ido} and the nonlocal approach 
to the cosmological constant problem \cite{Carroll, Oda1, Oda2}, though they are closely related to each other.
In the vacuum energy sequestering model, two gauge invariant variables of general relativity, in essence
the cosmological constant and the Planck mass scale, are promoted to dynamical variables, which play
a critical role in ensuring that the cosmological constant automatically cancels the radiative corrections 
from matters in the gravitational field equations. Meanwhile, in the nonlocal approach to the cosmological 
constant problem, a nonlocal constraint, which forces the total action to vanish identically, plays a role in
removing the cosmological constant from the Einstein equations. In the both models, operation of taking the
space-time average of some quantities brings nonlocal effects into the models \cite{Linde, Tseytlin, Nima} 
and consequently the effective
cosmological constant is expressed in terms of the space-time average of the trace of the energy-momentum
tensor (plus some additional terms in the latter model). 

In this way, Weinberg's no go theorem lays a cornerstone on nonlocal studies of the cosmological constant 
problem. However, this theorem is a purely classical statement based on field equations including classical
general relativity, so it is valuable to extend the classical theorem to a quantum mechanical one. Indeed, 
the cosmological constant problem stems from a clash between particle physics which sources the vacuum
energy density through quantum effects and gravity which responds to it classically. Moreover, in order to
describe this issue more accurately, it would be necessary to take account of quantum effects from graviton
loops, in other words, quantum gravity. In this article, we would like to present a purely quantum mechanical
proof of the Weinberg theorem within the framework of quantum gravity. 

In the proof by Weinberg, the general linear invariance $GL(4)$ plays a key role. However, it is known that this global 
invariance is broken spontaneously in quantum gravity. First of all, we wish to briefly review the manifestly 
covariant canonical formalism of quantum gravity \cite{Nakanishi1, Nakanishi-Ojima1} and 
account for why $GL(4)$ is broken spontaneously in quantum gravity \cite{Nakanishi-Ojima2}.  

The total action of quantum gravity is of form 
\begin{eqnarray}
S = \int d^4 x \ {\cal{L}},
\label{QG Action}
\end{eqnarray}
where the total Lagrangian density $ {\cal{L}}$ is defined as (for simplicity, we have put $8 \pi G_N = 1$ where 
$G_N$ is Newton's constant)
\begin{eqnarray}
{\cal{L}}  = \frac{1}{2} \sqrt{- g}  ( R - 2 \Lambda ) + \partial_\mu ( \sqrt{- g} g^{\mu\nu} ) b_\nu 
- i \sqrt{- g} g^{\mu\nu} \partial_\mu \bar c_\rho \partial_\nu c^\rho + {\cal{L}}_m.
\label{QG Lagr}
\end{eqnarray}
Here $g$ is the determinant of the metric tensor, $g = \det g_{\mu\nu}$, and $R$ and $\Lambda$ denote the scalar 
curvature and a bare cosmological constant, respectively. 
$b_\mu$ is an auxiliary field, and $c^\mu$ and $\bar c_\mu$ denote the FP ghosts and ${\cal{L}}_m$ denotes
the Lagrangian density for generic matter fields. In this action, as the gauge condition for diffeomorphisms, the 
following de Donder condition is chosen:
\begin{eqnarray}
\partial_\mu ( \sqrt{- g} g^{\mu\nu} ) = 0. 
\label{de Donder1}
\end{eqnarray}
Owing to the identity $\nabla_\mu ( \sqrt{- g} g^{\mu\nu} ) = 0$, Eq. (\ref{de Donder1}) can be rewritten as
\begin{eqnarray}
\Gamma^\mu_{\nu\rho}  g^{\nu\rho}  = 0, 
\label{de Donder2}
\end{eqnarray}
which is manifestly invariant under the general linear transformation $GL(4)$.

The action $S$ is not invariant under diffeomorphisms any longer, but it is still invariant under the BRST transformation 
and $GL(4)$ transformation. Actually, the $GL(4)$ generators have been found to be
\begin{eqnarray}
M^\mu \, _\nu = \int d^3 x  \sqrt{- g} g^{0 \rho}  \Bigl[ x^\mu  \partial_\rho b_\nu - \delta^\mu_\rho b_\nu
- i \bar c_\nu \partial_\rho c^\mu + i ( \partial_\rho \bar c_\nu ) c^\mu \Bigr].
\label{GL generator}
\end{eqnarray}
Using the canonical commutation relations, it is straightforward to calculate the following commutation relations 
\cite{Nakanishi-Ojima1, Nakanishi-Ojima2}: 
\begin{eqnarray}
&{}& [ M^\mu \, _\nu, M^\rho \, _\sigma ] = - i \delta^\mu_\sigma M^\rho \, _\nu + i \delta^\rho_\nu M^\mu \, _\sigma, 
\nonumber\\
&{}& [ g_{\rho\sigma}, M^\mu \, _\nu ] =  i x^\mu \partial_\nu g_{\rho\sigma} + i \delta^\mu_\rho g_{\nu\sigma}
+ i \delta^\mu_\sigma g_{\nu\rho}.
\label{CR}
\end{eqnarray}

If we assume that the translational invariance is not spontaneously broken, the vacuum expectation value of $g_{\mu\nu}$
is a flat Minkowski metric
\begin{eqnarray}
\langle 0| g_{\mu\nu} | 0 \rangle = \eta_{\mu\nu},
\label{VEV of g}
\end{eqnarray}
where $| 0 \rangle$ denotes the true vacuum state. The vacuum expectation value of the remaining fields, those are,
the auxiliary field $b_\mu$, the FP ghost $c^\mu$ and the FP antighost $\bar c_\mu$, is taken to be zero.  
Then, the latter commutation relation in Eq. (\ref{CR}), together with Eq. (\ref{VEV of g}), yields
\begin{eqnarray}
\langle 0| [ g_{\rho\sigma}, M^\mu \, _\nu ] | 0 \rangle  = i \delta^\mu_\rho \eta_{\nu\sigma}
+ i \delta^\mu_\sigma \eta_{\nu\rho}.
\label{SSB}
\end{eqnarray}
This equation clearly shows that the $GL(4)$ invariance is broken spontaneously. On the other hand, the Lorentz symmetry,
which is a subgroup of $GL(4)$, is exactly preserved since we can show that 
\begin{eqnarray}
\langle 0| [ g_{\rho\sigma}, J_{\mu\nu} ] | 0 \rangle  =  0,
\label{VEV Lorentz}
\end{eqnarray}
where the Lorentz generators $J_{\mu\nu}$ are defined as
\begin{eqnarray}
J_{\mu\nu} =  \eta_{\mu\rho} M^\rho \, _\nu - \eta_{\nu\rho} M^\rho \, _\mu.
\label{Lorentz}
\end{eqnarray}
As a result, the number of the spontaneous symmetry breakdown is equal to $16 - 6 = 10$, which precisely coincides with
the number of the dynamical degrees of freedom of the graviton. Thus, we can conclude that the graviton must be exactly massless 
owing to the Goldstone theorem \cite{Nakanishi-Ojima1, Nakanishi-Ojima2}.  Note that this proof is an exact proof 
without recourse to perturbation theory.  

Now we wish to present a quantum mechanical proof of the Weinberg theorem on the basis of the manifestly
covariant canonical formalism of quantum gravity.  Before doing so, let us ask two questions 
which are the key to our proof. First, recall that in the classical proof by Weinberg, the $GL(4)$ invariance plays a critical role, 
but as explained above,  $GL(4)$ is spontaneously broken in quantum gravity. Thus, the question to be asked first 
is what symmetry we can rely on in quantum gravity instead of $GL(4)$. This symmetry should be a global symmetry which is 
preserved exactly in quantum regime. The almost unique candidate for such a symmetry is nothing but the BRST symmetry 
which is a residual global symmetry of diffeomorphisms emerging after gauge fixing. 

Next, recall that the classical Lagrangian density and field equations also play a role in the Weinberg's proof. Thus, the second 
question which we should ask ourselves is which quantum mechanical quantity plays a similar role to the classical Lagrangian density. 
The answer is obvious again, namely, the effective action $\Gamma[\varphi]$, which can be obtained from the generating functional of
connected Green's functions, $W[J]$, via the Legendre transformation. Of course, the effective action $\Gamma[\varphi]$ is 
the generating functional of the 1PI (one-particle-irreducible) vertex functions $\Gamma^{(n)}$ and involves all information 
on radiative corrections in addition to classical action \cite{Peskin}. 

Armed with these ideas, we are now ready to present our proof. Let us note that the total action (\ref{QG Action}) is invariant 
under the following BRST transformation:
\begin{eqnarray}
\delta_B g_{\mu\nu} &=& - \partial_\mu c^\rho  g_{\rho\nu} - \partial_\nu c^\rho  g_{\rho\mu},  \quad
\delta_B c^\mu = 0,  \quad   \delta_B \bar c_\mu = i b_\mu,  \quad   \delta_B b_\mu = 0,
\nonumber\\ 
\delta_B \varphi_i &=& 0,  \quad \delta_B A_\mu = - \partial_\mu c^\rho A_\rho,  \quad
\delta_B x^\mu = c^\mu, 
\label{BRST}
\end{eqnarray}
where we have considered $N$ real scalar fields $\varphi_i ( i = 1, 2, \cdots, N )$ and a vector field $A_\mu$ as the matter fields. 
Let us recall that the conventional BRST transformation is given by
\begin{eqnarray}
\hat{\delta}_B g_{\mu\nu} &=& - c^\rho \partial_\rho g_{\mu\nu} - \partial_\mu c^\rho  g_{\rho\nu} - \partial_\nu c^\rho  g_{\rho\mu}
= - \nabla_\mu c_\nu - \nabla_\nu c_\mu, 
\nonumber\\
\hat{\delta}_B c^\mu &=& - c^\rho \partial_\rho c^\mu,   \quad
\hat{\delta}_B \bar c_\mu = i \hat{b}_\mu,  \quad   
\hat{\delta}_B \hat{b}_\mu = 0,  \quad
\hat{\delta}_B \varphi_i = - c^\rho \partial_\rho \varphi_i,   
\nonumber\\
\hat{\delta}_B A_\mu &=& - c^\rho \partial_\rho A_\mu - \partial_\mu c^\rho A_\rho,   \quad
\hat{\delta}_B x^\mu = 0. 
\label{BRST2}
\end{eqnarray}
It is known that the two types of nilpotent BRST transformations, (\ref{BRST}) and (\ref{BRST2}), are mathematically equivalent and 
they are simply related by the equation 
\begin{eqnarray}
\hat{\delta}_B \Phi = \delta_B \Phi - c^\rho \partial_\rho \Phi,
\label{BRST-relation}
\end{eqnarray}
with $\Phi \equiv \{ g_{\mu\nu}, \varphi_i, A_\mu, c^\mu,  \bar c_\mu, b_\mu, x^\mu \}$ and 
$\hat{b}_\mu \equiv b_\mu - i c^\rho \partial_\rho \bar c_\mu$ \cite{Nakanishi-Ojima1}, so we can use either at will.  
However,  in the case at hand, the former BRST transformation (\ref{BRST}) is more convenient than the latter one (\ref{BRST2}) since the Lagrangian
density transforms as a density under the BRST transformation (\ref{BRST}). We therefore use the former BRST transformation (\ref{BRST}) in this
article.   Under the BRST transformation  (\ref{BRST}), since $\sqrt{-g} d^4 x$ is the invariant volume, $\frac{1}{\sqrt{-g}} {\cal{L}}$ is BRST-invariant
whose fact can be verified by the explicit calculation. 

Next, to proceed in parallel with discussions on $GL(4)$ \cite{Padilla}, let us rewrite the BRST transformation of $g_{\mu\nu}$ as
\begin{eqnarray}
\delta_B g_{\mu\nu} = - \partial_\mu c^\rho  g_{\rho\nu} - \partial_\nu c^\rho  g_{\rho\mu} = \delta M_{\mu\nu} + \delta M_{\nu\mu}.
\label{g-BRST}
\end{eqnarray}
Since the BRST transformation (\ref{BRST}) is the residual transformation of diffeomorphisms like $GL(4)$, it is convenient 
to express the BRST transformation in terms of the $GL(4)$-like expression as (precisely speaking, we should put
the Grassmann-odd parameter $\lambda$ in front of $\delta M^\mu \, _\nu$, but we omit it since this parameter is
irrelevant for later argument)
\begin{eqnarray}
M^\mu \, _\nu = \delta^\mu_\nu + \delta M^\mu \, _\nu.
\label{GL-BRST}
\end{eqnarray}
At this stage, let us consider an integrand of the effective action, $\tilde \Gamma$ which is defined as
\begin{eqnarray}
\Gamma = \int d^4 x \ \tilde \Gamma,
\label{tilde EA}
\end{eqnarray}
where $\Gamma$ is the conventional effective action which is invariant under the BRST transformation (\ref{BRST}).  
(It is assumed that we can obtain a BRST-invariant effective action $\Gamma$ by following a recipe explained in Ref. \cite{Parker}.)
Since $\tilde \Gamma$ is a density quantity under the BRST transformation, it should be transformed as
\begin{eqnarray}
\tilde \Gamma \rightarrow \tilde \Gamma^\prime = (\det M) \tilde \Gamma.
\label{BRST tilde EA}
\end{eqnarray}
Then, in the infinitesimal form, the BRST transformation of $\tilde \Gamma$ reads
\begin{eqnarray}
\delta_B \tilde \Gamma = \tilde \Gamma^\prime - \tilde \Gamma \approx (Tr \delta M) \tilde \Gamma
= - (\partial_\rho c^\rho) \tilde \Gamma.
\label{inf-BRST tilde EA}
\end{eqnarray}
In fact, using Eq. (\ref{inf-BRST tilde EA}) and $\delta_B \sqrt{-g} = - \sqrt{-g} \partial_\rho c^\rho$, it is easy to show that 
$\frac{1}{\sqrt{-g}} \tilde \Gamma$ is invariant under the BRST transformation, thereby meaning that the effective action 
$\Gamma$ in Eq. (\ref{tilde EA}) is BRST-invariant as required.

Now let us assume that the translational invariance is not broken spontaneously, which indicates Eq. (\ref{VEV of g}) and
\begin{eqnarray}
\langle 0| \varphi_i | 0 \rangle = \varphi_i^{(0)},  \quad \langle 0| A_\mu | 0 \rangle = 0,
\label{VEV of phi}
\end{eqnarray}
where $\varphi_i^{(0)}$ are constant modes independent of the space-time coordinates. Since we have a constant vacuum 
solution Eq. (\ref{VEV of g}) and Eq. (\ref{VEV of phi}) (of course, $b_\mu$, $c^\mu$ and $\bar c_\mu$ have a vanishing 
vacuum expectation value),  we can infer the following relation
\begin{eqnarray}
\delta_B \tilde \Gamma = \frac{\partial \tilde \Gamma}{\partial \varphi_i} \delta_B \varphi_i 
+ \frac{\partial \tilde \Gamma}{\partial A_\mu} \delta_B A_\mu 
+ \frac{\partial \tilde \Gamma}{\partial g_{\mu\nu}} \delta_B g_{\mu\nu} 
+ \frac{\partial \tilde \Gamma}{\partial c^\mu} \delta_B c^\mu
+ \frac{\partial \tilde \Gamma}{\partial \bar c_\mu} \delta_B \bar c_\mu
+ \frac{\partial \tilde \Gamma}{\partial b_\mu} \delta_B b_\mu. 
\label{Relation}
\end{eqnarray}
Meanwhile, for the translational invariant fields, the quantum field equations, or the generalized Euler-Lagrange equations, 
which involve all quantum effects, are given by
\begin{eqnarray}
\frac{\partial \tilde \Gamma}{\partial \varphi_i} = \frac{\partial \tilde \Gamma}{\partial A_\mu} 
= \frac{\partial \tilde \Gamma}{\partial g_{\mu\nu}} = \frac{\partial \tilde \Gamma}{\partial c^\mu}
= \frac{\partial \tilde \Gamma}{\partial \bar c_\mu} = \frac{\partial \tilde \Gamma}{\partial b_\mu} 
= 0.
\label{q-eq}
\end{eqnarray}

With the field equations except $\frac{\partial \tilde \Gamma}{\partial g_{\mu\nu}} = 0$ in Eq.  (\ref{q-eq}), together with 
Eq. (\ref{inf-BRST tilde EA}), Eq.  (\ref{Relation}) gives us a relation (In fact, only quantum field equations
$\frac{\partial \tilde \Gamma}{\partial A_\mu} = \frac{\partial \tilde \Gamma}{\partial \bar c_\mu} = 0$ 
are needed because of the BRST transformation (\ref{BRST}).)
\begin{eqnarray}
\delta_B \tilde \Gamma = \frac{\partial \tilde \Gamma}{\partial g_{\mu\nu}} \delta_B g_{\mu\nu}
= \frac{\partial \tilde \Gamma}{\partial g_{\mu\nu}} (- 2 \partial_\mu c^\rho g_{\rho\nu})
= - ( \partial_\rho c^\rho ) \tilde \Gamma,
\label{New Relation}
\end{eqnarray}
which provides us with the equation
\begin{eqnarray}
\frac{\partial \tilde \Gamma}{\partial g_{\mu\nu}} - \frac{1}{2} g^{\mu\nu} \tilde \Gamma = 0.
\label{Final eq}
\end{eqnarray}
This equation can be easily solved to be
\begin{eqnarray}
\tilde \Gamma = \sqrt{-g} V(\varphi_i),
\label{Final ans}
\end{eqnarray}
where $V(\varphi_i)$ is a certain function of only $\varphi_i$.  (Since we assume the translational invariance
in this article, $V$ is nothing but the effective potential.)
Finally, imposing $\frac{\partial \tilde \Gamma}{\partial g_{\mu\nu}} = 0$, we obtain 
\begin{eqnarray}
V(\varphi_i) = 0,
\label{V=0}
\end{eqnarray}
which corresponds to fine tuning of the cosmological constant. In this way, we have succeeded in proving
a quantum mechanical generalization of the Weinberg no go theorem.

To summarize, in this article, based on the manifestly covariant canonical formalism of quantum gravity \cite{Nakanishi1, Nakanishi-Ojima1}, 
we have proved that Weinberg's no go theorem, which is originally a statement in classical gravity, is valid even 
in quantum gravity. Even if we make use of a specific canonical formalism of quantum gravity, our proof is quite general 
in that we rely on only the BRST invariance of the effective action of quantum gravity, so we believe that the Weinberg theorem 
holds in a general formalism of quantum gravity. 
 
To tell the truth, a hidden motivation behind this article has lain in a primitive question of the present author: If some gravitational
theory does not possess the $GL(4)$ symmetry that survives as a vestige of diffeomorphisms when the fields are restricted to 
be constants, can such a gravitational theory evade Weinberg's no go theorem?  As such a gravitational theory, we have had
in mind the transverse Weyl gravity which is not invariant under $GL(4)$ but is equivalent to general relativity at least at the classical
level \cite{Oda3}.  The result obtained in this article also answers this question in a negative way: The transverse Weyl gravity cannot 
evade the Weinberg theorem since although there is no the $GL(4)$ symmetry there is a BRST symmetry in this gravitational theory 
as well.

\begin{flushleft}
{\bf Acknowledgements}
\end{flushleft}
This work is supported in part by the Grant-in-Aid for Scientific 
Research (C) No. 16K05327 from the Japan Ministry of Education, Culture, 
Sports, Science and Technology.


\end{document}